\documentclass[aps,floatfix,twocolumn,a4paper,showpacs, nofootinbib,superscriptaddress,10pt]{revtex4}
\usepackage{graphicx,float}\usepackage{graphicx,float}
\usepackage[all]{xy}
\usepackage{amsmath,upgreek}
\usepackage{amssymb}
\usepackage{color}
\usepackage{epsfig}		
\usepackage{graphicx,epstopdf}
\usepackage{subfigure}
\usepackage{pdfpages}
\usepackage[colorlinks,hyperindex]{hyperref}

\hypersetup{
colorlinks,
citecolor=blue,
linkcolor=black,
urlcolor=black,
}

\newcommand{\bes}{\begin{subequations}}
\newcommand{\ees}{\end{subequations}}
\def\ben{\begin{eqnarray}}
\def\een{\end{eqnarray}}
\def\be{\begin{equation}}
\def\ee{\end{equation}}

\begin{document}

\title{Conditional entropy of glueball states }
\author{Alex E. Bernardini} 
\affiliation{Departamento de F\'isica, Universidade Federal de S\~ao Carlos,
PO Box 676, 13565-905, S\~ao Carlos, SP, Brazil}
\email{alexeb@ufscar.br}
\author{Nelson R. F. Braga}
\affiliation{Instituto de F\'isica, Universidade Federal do Rio de Janeiro,
Caixa Postal 68528, RJ 21941-972 - Brazil}
\email{braga@if.ufrj.br}
\author{Rold\~ao da Rocha}
\affiliation{CMCC, Universidade Federal do ABC, UFABC, 09210-580, Santo Andr\'e, Brazil}
\email{roldao.rocha@ufabc.edu.br}

\begin{abstract}
The  conditional entropy of glueball states is calculated using a holographic description. Glueball states are represented by a supergravity dual picture, consisting of a 5-dimensional graviton-dilaton 
action of a dynamical holographic AdS/QCD model. The  conditional entropy is studied as a function of the glueball spin and of the mass, providing  information about the stability of the glueball states. 
\end{abstract}
\maketitle

\section{Introduction}

AdS/QCD models, inspired in the AdS/CFT correspondence 
\cite{Maldacena:1997re,Gubser:1998bc,Witten:1998qj}, provide an important phenomenological tool for describing hadronic properties in 
the low-energy regime, where QCD is non perturbative.  The hadronic states are represented in the dual supergravity picture by normalizable solutions of fields that live in a five-dimensional anti-de Sitter (AdS$_5$) space, endowed with a
hard \cite{Polchinski:2001tt,BoschiFilho:2002ta,BoschiFilho:2002vd} or soft \cite{Karch:2006pv}
infrared (IR) cut off.  The cut off in AdS space breaks conformal invariance, introducing a mass parameter  in the models, that sets the scale for the mass spectra of hadrons.

Glueballs are  bound states of gluons, that are expected to  appear in high energy physical processes, as a consequence of the self-coupling of gluons in QCD. Conclusive experimental data, about these type of particles, still lack. 
Lattice QCD provides an important  tool to calculate glueball masses (see for example 
\cite{Morningstar:1999rf,Lucini:2001ej, Chen:2005mg}). On the other hand, the decay process of glueballs (and other hadrons) is in general  difficult to describe.  One of the problems faced is that radially excited states, that have the same quantum numbers, get mixed in lattice imaginary time numerical simulations.

The conditional entropy is an interesting tool for investigating the configurational stability underlying physical systems. 
It has recently been shown, for the case of mesons, that  the conditional entropy measures the relative occurrence of the physical states \cite{Bernardini:2016hvx}, suggesting that the entropy indeed provides information about the relative stability of states. 
Here we will apply the lattice approach of Shannon entropy \cite{Gleiser:2013mga,Gleiser:2014ipa} and its statistical mechanics underlying structure, described in   \cite{Bernardini:2016hvx}, to the glueball case.
We shall represent the glueballs using a recent model proposed in refs. \cite{Capossoli:2016ydo,Capossoli:2015ywa}, that is a modified soft-wall model and provides nice fits of glueball masses for even and odd spins.  We will develop a procedure that leads to a relation between the glueballs spins -- and the glueballs masses --  and the associated  conditional entropy.   This analysis can shed some light into the relative stability of the different glueball states. 

The so-called information entropy setup is related to the irresolution of information in a physical system \cite{Gleiser:2014ipa,Correa:2016pgr}. Besides, the conditional entropy can extend the Shannon information entropy \cite{shannon}  to some continuum limit of modes, that comprise the physical system.   Recently, the modal fractions -- in information entropy theory -- were defined as the ratio between collective coordinates and the structure factor  -- in the thermodynamical entropy setup -- further providing the statistical mechanical analogue of the conditional entropy setup \cite{Bernardini:2016hvx}.

This work is organized as follows: in Sect. II, the anomalous dynamical AdS/QCD holographic model is introduced by a dilaton-graviton bulk action, with a subsequent scalar glueball action. A beta function with an IR fixed point, at finite coupling, is then used in the model. The dimension of the operators in the $\mathcal{N} = 4$  CFT is used for defining the 5-dimensional glueball mass, and hence, the 4-dimensional glueball mass, as a function of the glueball spin and the beta function.  Employing the collective coordinates 
and the structure factor, calculated upon the energy density of the system,  the thermodynamical entropy is a foundation to compute the conditional entropy associated with glueball states. Hence the information entropy, and the stability of glueballs, are quantitatively studied, for different values of the model parameters.  
Our concluding remarks are presented  in Sect. III.

\section{Conditional entropy and glueball stability}
The energy density of the bulk modes, associated with the glueball states,  is a relevant tool for the information entropy  analysis of the glueballs stability, in the AdS/QCD framework. Glueballs are predicted by QCD and are modeled using lattice gauge
theory. The ground state is the scalar glueball  $0^{\tiny{++}}$ that is expected, from lattice computation, to have a mass 1.6 to 1.7 GeV \cite{Agashe:2014kda}.  The
search for this state has been, and still is, in the center of vivid activity in the framework of low-energy QCD. This
state is also important because it is related to two basic phenomena of QCD:  the generation of the gluon condensate and also the  anomalous breaking  of dilatation
invariance  \cite{Janowski:2014ppa}.

 Recently, a new holographic model for calculating glueball masses appeared in Ref. \cite{Capossoli:2016ydo}.
 It consists of a modification of the soft-wall model \cite{Karch:2006pv}, that is analytically solvable and provides the masses for the high spin states.  In this framework, the $5$-dimensional action for the graviton-dilaton coupling reads, in the Einstein frame, 
\begin{equation}\label{einsteinframe}
S =  \int \!d^5 x \sqrt{-g} \left({\rm R} -\frac{4}{3}g_{MN} \partial^M \phi \partial^N \phi - V(\phi)\right),
\end{equation} where   $\phi=\phi(z)$ denotes the dilaton field,  $V(\phi)$ stands for the dilatonic potential
and the conformal metric $g_{MN} dx^M dx^N$ has the form
\begin{equation}\label{g}
ds^2 =  
e^{2A(z)}(g_{\mu \nu}dx^\mu dx^\nu+dz^2),
\end{equation}
\noindent with $\mu, \nu=  0,1,2,3$  and  $g_{\mu \nu}$ denotes the  Minkowski metric; whereas the 5-dimensional AdS indices attain the values $M,N,Q,R,S = 0,1,2,3,4$. The 5-dimensional metric determinant is denoted by $g$ and the Einstein-Hilbert part of the action in (\ref{einsteinframe}) regards the scalar curvature ${\rm R}$. 
  Hereon, normalized units $16 \pi G_5=1$ shall be adopted, where  $G_5$ is the Newton 5-dimensional coupling constant.  The equations of motion read \cite{German:2012rv,Dutra:2013jea,Bazeia:2013usa}
 \begin{eqnarray}\label{mov1}
 \!\!\!\!\!\!\!G_{RS} \! -\! \frac{1}{2}g_{RS}\!\left[\frac{4}{3}\!\left( 2\partial^R \phi \partial^S \phi \!-\!\partial_Q\phi \partial^Q\phi\right)\!-\! V(\phi) \!\right] \!\!&\!=\!& 0,\\\label{mov2}
\!\!\!\!\!\!\! \frac{3 \sqrt{g}}{8}\frac{dV(\phi)}{d\phi}-g_{RS} \partial^R (\sqrt{g} \partial^S \phi)  \! &\!=\!& 0,
\end{eqnarray}
\noindent where $G_{RS}$ is the Einstein tensor.

Using the conformal metric given by Eq. (\ref{g}), the equations of the motion (\ref{mov1}, \ref{mov2})  yield, by  denoting $B'(z)=dB/dz$, for any quantity $B$:
\begin{eqnarray}
 -\phi'' - 3 \phi'\mathring{A}' + \frac{3}{8}e^{2\mathring{A}} \frac{dV(\phi)}{d\phi} &=& 0, \label{mov3}\\\label{mov4}
 \mathring{A}''+ \frac{4}{9}\phi'^2 - \mathring{A}'^2  &=& 0,
\end{eqnarray}
\noindent where \cite{Capossoli:2015ywa,BoschiFilho:2012xr}
\begin{equation}\label{warp}
 \mathring{A}(z) = - \frac{2}{3}\phi(z)+A(z) \;.
\end{equation}
\noindent Solving Eqs. (\ref{mov3}) and (\ref{mov4}) for the quadratic dilaton background, 
\begin{eqnarray}\label{qadr}
\phi(z)=kz^2,
\end{eqnarray} it yields expressions for the warp factor and the potential, respectively:
\begin{eqnarray}\label{eq444}
 \!\!\!\!\!\!\!\!\mathring{A}(z) &=&- \ln{\left[_0F_1\left(\frac 54, \frac{\phi^2}{9}\right)\frac{z}{R} \right]},\label{fjk}\\
 \label{potencial}
 \!\!\!\!\!\!\!\!\!\! R^2V(\phi)\! &\!\!=\!\!&\! \! \frac{16\phi^2}{3} {}_0F_1^2\left(\frac 54, \frac{\phi^2}{9}\right)\!-\!{12 ~ _0F_1^2\left(\frac14, \frac{\phi^2}{9}\right)},
\end{eqnarray}
where $R$ denotes the AdS radius and $_0F_1$ is a confluent hypergeometric limit function\footnote{It is related to the Bessel functions \\$J_\alpha(x) = \frac{(x/2)^\alpha}{\Gamma(\alpha+1)}{}_0F_1\left(\;\cdot\;,\alpha+1,-\frac{1}{4}x^2\right)$.}.  
Using Eqs. (\ref{warp}) and (\ref{eq444}) yields 
\begin{equation}\label{warp_2}
 A(z) =   \frac{2}{3}\phi(z) - \ln{\left[\frac{z}{R}{}_0F_1\left(\frac 54, \frac{\phi^2}{9}\right) \right]}\;.
\end{equation}
It means that the metric in Eq. \eqref{g} in this dynamical model is an asymptotically AdS$_5$ metric,   in the ultraviolet (UV) limit \cite{Capossoli:2015ywa,BoschiFilho:2012xr,forkel}.

Now, the 5-dimensional action for the scalar glueball, represented by the field $\mathfrak{G}$,  has the following form  \cite{Capossoli:2016ydo,Capossoli:2015ywa,BoschiFilho:2012xr}:
\begin{equation}\label{glue}
S = \frac{1}{2}\int d^5 x \sqrt{-g} \;  e^{-\phi(z)}\left[g_{RS} \partial^R \mathfrak{G}\partial^S \mathfrak{G} + M^2_{5} \mathfrak{G}^2\right]\;,
\end{equation}
\noindent and its equations of motion, using the metric (\ref{g}), are expressed as: 
\begin{equation}\label{mfk}
e^{-\phi(z)+5 A(z)} M^2_{5} \mathfrak{G}-g_{RS}\partial^R[e^{- \phi(z)+3A(z) } \partial^S \mathfrak{G}]  = 0\;.
\end{equation}
\noindent Considering the ansatz $
\mathfrak{G}(x^{\mu},z) \equiv \mathfrak{G}(z) e^{i p_{\mu} x^{\mu}},$ where
\begin{equation}\label{ansatz1}
\mathfrak{G}(z) = \psi (z) \exp\left({\frac{k^2z^4 - 3A(z)}{2}}\right),
\end{equation}
 and using  the quadratic dilaton of eq. (\ref{qadr}),  one finds 
\begin{eqnarray}\label{duas}
\!\! \psi'' \!\!\!\!\!\!\!\!\!&&\!\!\!-\! \left[ k^2 z^2 \!+\! \frac{1}{z^2}\!\!\left(\!\frac{15}{4}\!+\!{M_{5}^2R^2}  e^{\frac43{kz^2}}\! {}_0F_1^{-2}\left(\frac14,\! \frac{k^2z^4}{9}\right)\!\!\right) \! -\! 2k \right] \!\psi\nonumber\\\!\!&\!=&  p^\mu p_\mu \psi \;. 
\end{eqnarray}
where, for simplicity,  we denote $\psi(z)$ by $\psi$.  
A similar Schr\"odinger-like equation was solved numerically in \cite{Li:2013oda}. The masses found for the scalar glueball and its radial (spin 0) excitations are  compatible with those obtained by lattice QCD. Up to linear order in $k$, and relating $p^\mu p_\mu$ to the 4-dimensional glueball states masses $m_n^2$, it yields \cite{Capossoli:2015ywa}:
\begin{equation}\label{4dm}
m_n^2=\left(4n+ \frac43M_5^2R^2+2(M_5^2R^2+4)^{1/2}\right)k,\quad \,
\end{equation}
where $n = 0, 1 , 2 , ...\,$. It is worth to mention that for the lightest scalar glueball state, corresponding to the spin $0^{\tiny{++}}$, that is dual to the bulk fields of zero mass, $M_5^2 = 0$, it yields $m_n^2 =  (4+4n)k$ \cite{Capossoli:2015ywa}.

The energy density associated with the glueball states immediately reads from the action in Eq. (\ref{glue}), by taking into account Eq. (\ref{ansatz1}) \cite{bloch}:
\begin{eqnarray}\label{eenergy}
\rho(z)= T^{00}(z)=\frac12\,e^{2A(z)}\!\left[\left(\mathfrak{G}'(z)\right)^2 \!+\! M^2_{5} \mathfrak{G}^2(z)\right].\end{eqnarray}

In order to take into account dynamical corrections as well as the anomalous dimension effects, glueball states have full dimension as a function of spin $J$ \cite{Capossoli:2015ywa}:
\begin{eqnarray}\label{evenodd}
\Delta(J) = 4 \! +\! 2\left[1\!+\!(-1)^{J+1}\right] J \!-\! \frac{2}{\lambda} \beta(\lambda) \!+\! \beta'(\lambda),
\end{eqnarray} where the spin $J=0,1,2,\ldots,$ shall thus define the even and odd glueball states. 
 It is worth to mention that  this expression comes from 
the correspondence between supergravity on AdS$_5 \times S^5$ and chiral fields in $\mathcal{N}=4$ (super)conformal theory in 4 dimensions \cite{Maldacena:1997re}. In this setup, the mass of a 0-form on AdS$_5$ is related to the dimension $\Delta$ in Eq. (\ref{evenodd}), of a 4-form operator, in the CFT, by the expression  \cite{ooguri} \begin{equation}\label{m55}
 M^2_5R^2 = \Delta (\Delta - 4).\end{equation}
It means that the full dimension $\Delta$ in Eqs. (\ref{evenodd}) gives the expression for the  bulk glueball mass $M_5$ \cite{Capossoli:2015ywa,BoschiFilho:2012xr}.

In order to describe even and odd spin glueball states, one can replace eq. (\ref{evenodd})  into the Schr\"odinger-like equation obtained from the dynamical soft-wall model, eq. (\ref{duas}),  and solve it numerically, for glueball states. 
Following ref. \cite{BoschiFilho:2012xr}, one chooses a beta function, with a finite coupling IR fixed point: 
\begin{equation}\label{betadef}
\beta(\lambda) \propto \lambda^2 \left(\frac{\lambda}{\lambda_{\star}}-1\right)\,,\,\qquad \lambda_{\star} > 0\;.
\end{equation}
 In ref. \cite{BoschiFilho:2012xr}, masses of glueball states  with even and odd spins were calculated for different values of the model parameters  $k$, for   $\lambda_{\star} = 350$. Table I shows results of \cite{BoschiFilho:2012xr,Capossoli:2016ydo}, according to eq. (\ref{4dm}).
\begin{widetext}
\begin{center}
\begin{table}[!h]
\centering
\begin{tabular}{||c||c||c|c|c|c|c|c||c|c|c|c|c|c||}
\hline\hline  & \multicolumn{1}{|c||}{}
&  \multicolumn{12}{c|}{Glueball states (odd and even spins) $J^{PC}$}  \\  
\cline{1-14}
Data & $k$ (GeV$^{2}$)&  $1^{\tiny{--}}$ & $3^{\tiny{--}} $ & $5^{\tiny{--}}$ & $7^{\tiny{--}}$ & $9^{\tiny{--}}$ & $11^{\tiny{--}}$ & $0^{\tiny{++}}$ & $2^{\tiny{++}} $ & $4^{\tiny{++}}$ & $6^{\tiny{++}}$ & $8^{\tiny{++}}$ & $10^{\tiny{++}}$    \\
\hline \hline                                                        
 I& $ 0.04$  & 2.99 & 3.89 & 4.75  &  5.63  & 6.49  & 7.33& 1.57 & 2.53 & 3.44 &  4.33  & 5.20  & 6.04  \\ \hline                                               
 II& $ 0.09$ & 3.43 & 4.56 & 5.71 &  6.83  & 7.92 & 9.05& 1.63 & 2.85  & 4.01 &  5.13  & 6.27  &  7.38      \\ \hline
 III& $ 0.16$  & 4.03 & 5.51 & 6.94 & 8.41  & 9.85   & 10.01  & 1.70 & 3.27 & 4.75 & 6.24  & 7.68  & 9.13   \\ \hline  
\hline
 \end{tabular} 
\caption{Glueball states  masses {\rm (GeV)}, for $C=P=\pm 1$, as a function of the glueball spin $J$, in the anomalous dynamical soft-wall model setup. The IR fixed point $\lambda_{\star}$ is 350 \cite{BoschiFilho:2012xr}. }
\label{t19}
\end{table}
\end{center}
\end{widetext}
It is worth to mention that  the results of Table I are in agreement with  glueball mass spectrum models, that predict  a maximum value of $1.7\pm0.1$ GeV for the mass of the ground state \cite{Agashe:2014kda}.
 The states $f_0$(1500)  or, alternatively, the $f_0$(1710), have been proposed as candidates for the scalar glueball  \cite{amsler,Janowski:2014ppa}. 

Now, the conditional entropy can be employed as the lattice approach of  Shannon information entropy, that was shown in \cite{Bernardini:2016hvx} to have underlying  statistical mechanics grounds. 
The entropic information, realized by the  conditional entropy, was used in the lattice to study physical systems \cite{Gleiser:2013mga,Gleiser:2014ipa}. We can use it  to study the stability of glueball state configurations, within the AdS/QCD setup. In fact, any  physical system  has the classical field configuration to be the one corresponding either to a critical point of the action, in the classical field theory setup, or to a critical point of the effective action, in a semiclassical approximation of a quantum theory. Furthermore, any physical system has the conditional entropy critical points 
corresponding to the most stable configurations, from the information entropy point of view \cite{Correa:2015lla}. 
States of higher conditional entropy either request a higher amount of energy to be created, or are more seldom detected (or observed) than their counterparts that present configurational stability, or both \cite{Bernardini:2016hvx}. 

The conditional entropy generalizes the information entropy for density functions that are naturally spatially localized, as the Fourier transform of the energy density function related to the 
physical setup. 
The information entropy was originally defined, for a system with $n$ modes, by $
S_c = -{{\sum_{j=1}^N}}\;h_j\ln(h_j),$
where  $\{h_j\}$ is a set of probability density functions ruling the physical system  \cite{shannon}.  The conditional entropy, hence, has critical points that define  configurations that are stable and that correspond to the best compression of information  in the system. More than a single stable configuration can appear, as in the case of oscillating configurations for the evolution of domain walls. In this case, phase transitions occur by the decay of the false vacuum \cite{Correa:2015rka}. 
Moreover, the conditional entropy can be analogous to the thermodynamical entropy,  being also potentially related to entanglement entropy  \cite{queiroz}.

 To implement the conditional entropy for  glueball states we use, as before, $z$ as the usual  bulk coordinate of AdS space and write the corresponding Fourier transform 
 of the energy density as:
\begin{equation}
\uprho(\upomega)=(2\pi)^{-1/2}\int_{-\infty}^{+\infty} \!\!\!\rho(z)\,e^{i\upomega z} dz\,.
\label{collectivecoordinates}
\end{equation}
This can be thought as a continuum limit of the well-known collective coordinates in statistical mechanics, $\rho(z)=\sum_{j=1}^N \uprho(\upomega_j)\exp\left({-i\upomega_j z}\right)$.    The structure factor, $
s_N = \frac1N\sum_{j=1}^N\;\langle\;\uprho(\upomega_j)\uprho^*(\upomega_j)\;\rangle\,,$ normalizes the correlation of collective coordinates, as  
\begin{equation}
f(\upomega_N)=\frac{1}{N\,s_N}{\langle\; \uprho(\upomega_N)\uprho^*(\upomega_N) \;\rangle}.\label{collective1}
\end{equation} The structure factor  measures energy density fluctuations and, hence, also the system behavior so as to approach  homogenization. By taking the $N \to \infty$ limit, and regarding  Eq. (\ref{collectivecoordinates}), the structure factor is then used, 
to yield the modal fraction to be defined as the correlation of collective coordinates-to-structure factor ratio:
 \begin{equation}
 f(\upomega)\equiv \lim_{N\to\infty}f(\upomega_N)=\frac{\big\langle\;\left\vert \uprho(\upomega)\right\vert ^{2}\big\rangle}{{\displaystyle{\lim_{N\to\infty}\int_{-N}^{N}}}
 \big\langle\;\left\vert \uprho(\upomega)\right\vert ^{2}\big\rangle \,d\upomega}\,\,. \label{collective}
 \end{equation}
 The lattice approach of the conditional entropy is denoted by \cite{Gleiser:2013mga,Gleiser:2014ipa}  
\begin{equation}
S_c[f]\;{=} -\lim_{N\to\infty}\int_{-N}^{N} \ln\left[f(\upomega)\right]\,f(\upomega)\,d\upomega \label{conditional}
\end{equation} 

Now, the anomalous dynamical soft-wall model is employed, 
to derive the relationship between the conditional entropy and 
the glueball state spins or, equivalently, the glueball state 4-dimensional masses. In fact, the energy density in Eq. (\ref{eenergy}) is appropriate for computing the conditional entropy, since they are spatially localized functions,  
encoded in the $T^{00}$ component. By using Eqs. (\ref{collective}) and (\ref{conditional}), that define the conditional entropy, together with Eq. (\ref{eenergy}) -- that takes into account Eqs. (\ref{qadr}) and (\ref{potencial}) for the warp factor and the dilaton potential, respectively -- the profiles for the conditional entropy, as a function of the glueballs spins, are then obtained.   

\begin{figure}[H]
\centering\includegraphics[width=7.9cm]{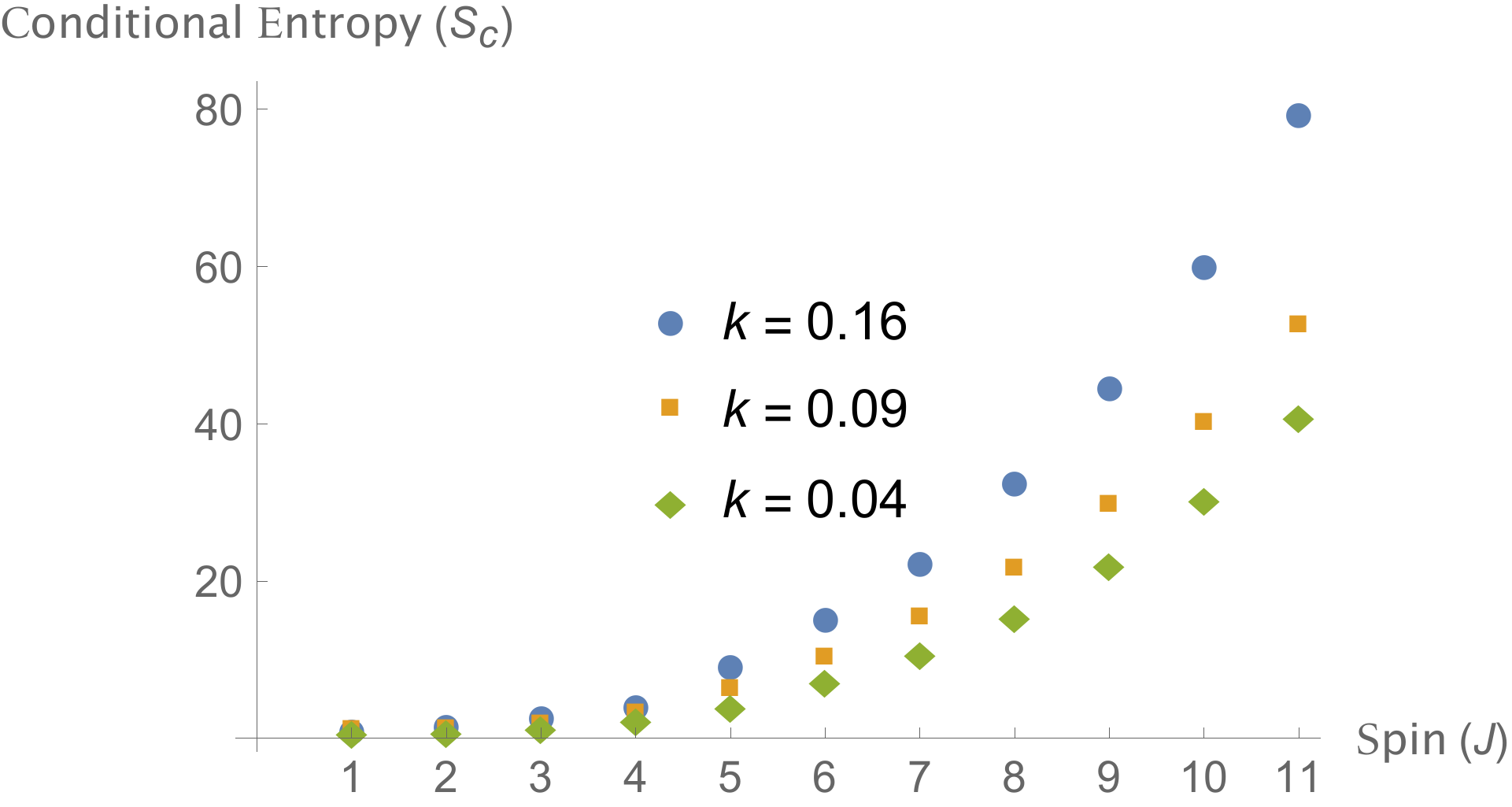}
\caption{Conditional entropy for $k=0.04$ (line 1, Table I), and $k=0.09$ (line 2, Table I), and $k=0.16$ (line 3, Table I) as functions of the spin.}
\label{f1}
\end{figure}

 To compute the conditional entropy for the energy density (\ref{eenergy}), it is worth to observe that when the UV limit is taken into account, the metric \eqref{g} is asymptotically AdS. The conditional entropy (\ref{conditional}) is calculated from the modal fraction (\ref{collectivecoordinates}), when Eqs. (\ref{m55}) and (\ref{betadef}) are regarded. The numerical results for three different values of $k$, as functions of the spin, are depicted in Fig. \ref{f1}. 
\begin{figure}[H]
\includegraphics[width=8.45cm]{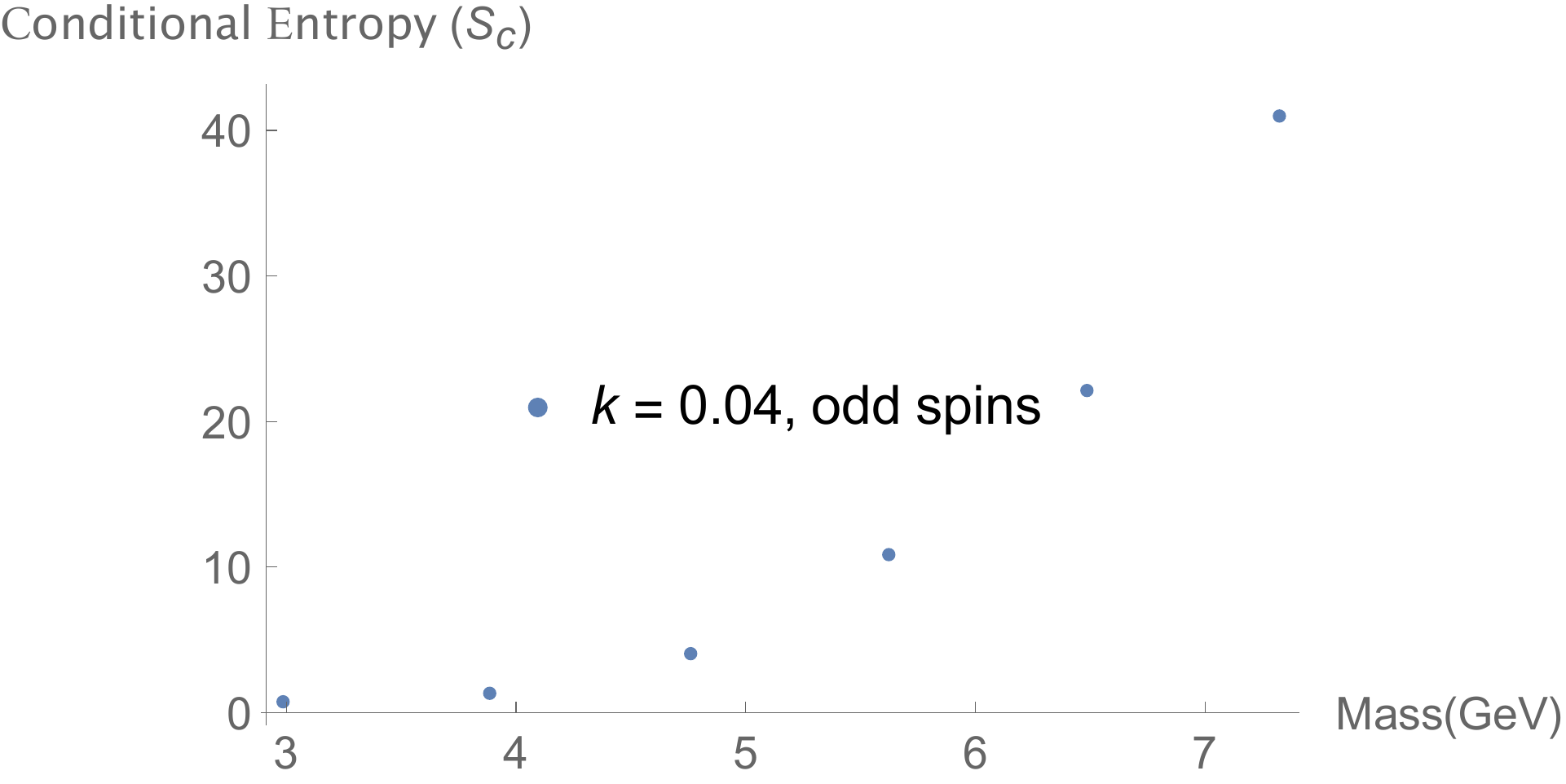}
\includegraphics[width=8.45cm]{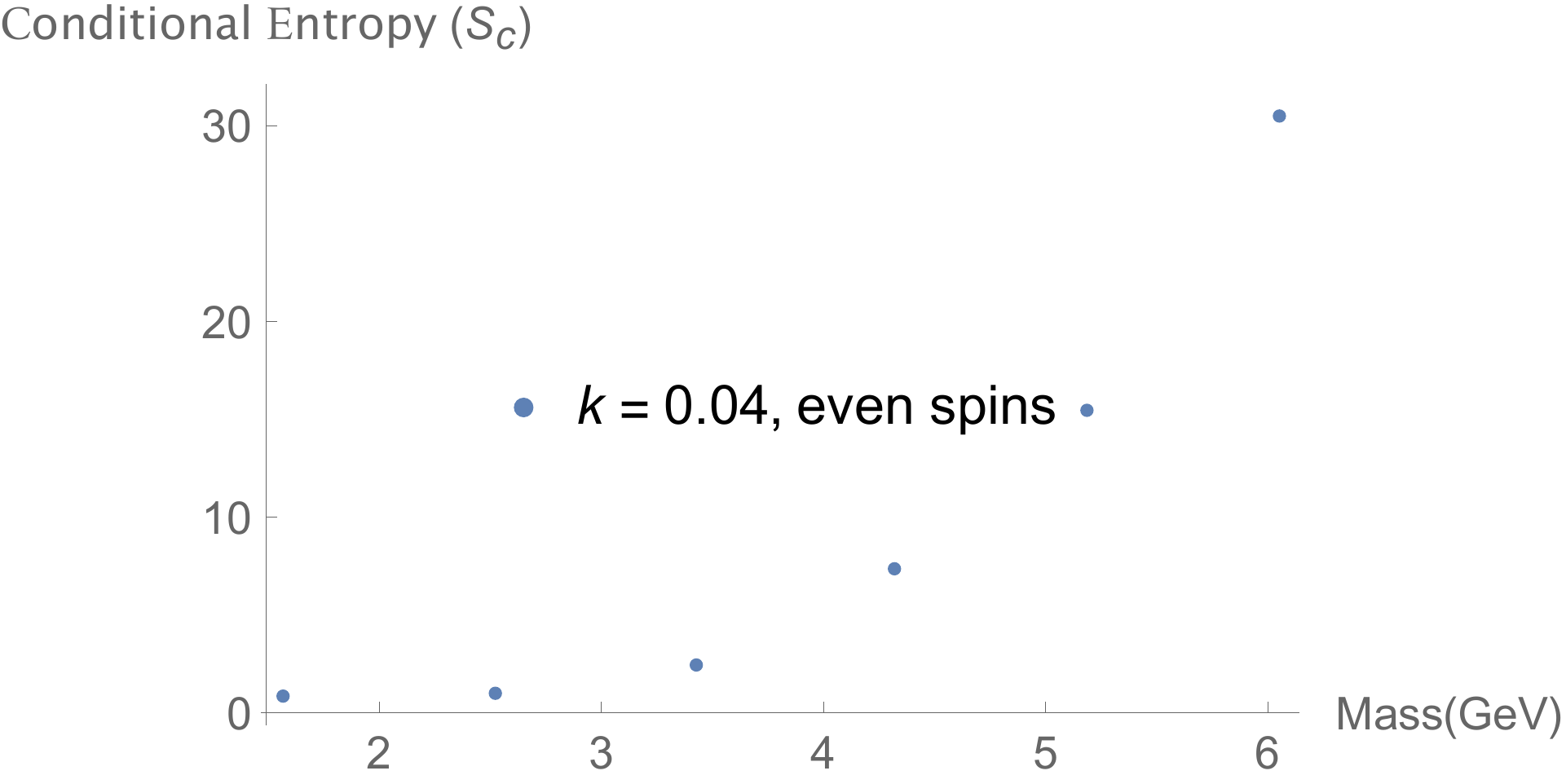}
\label{f4}
\caption{Conditional entropy of glueball states  for $k=0.04$ (line 1 in Table I)  for odd spin (top panel) and even spin (bottom panel) as a function of the glueball mass.}
\end{figure}
\begin{figure}[H]
\includegraphics[width=8.45cm]{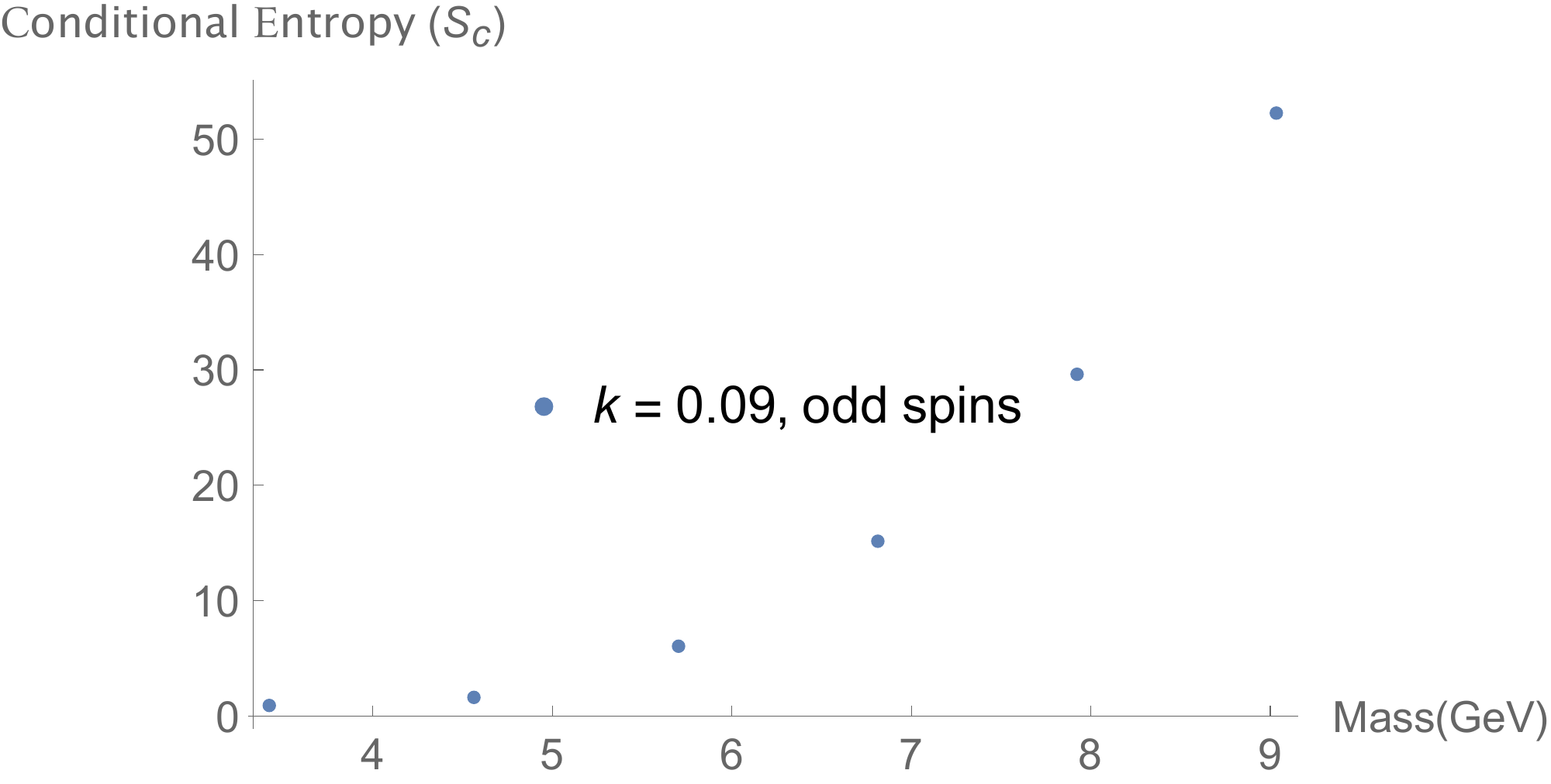}
\includegraphics[width=8.45cm]{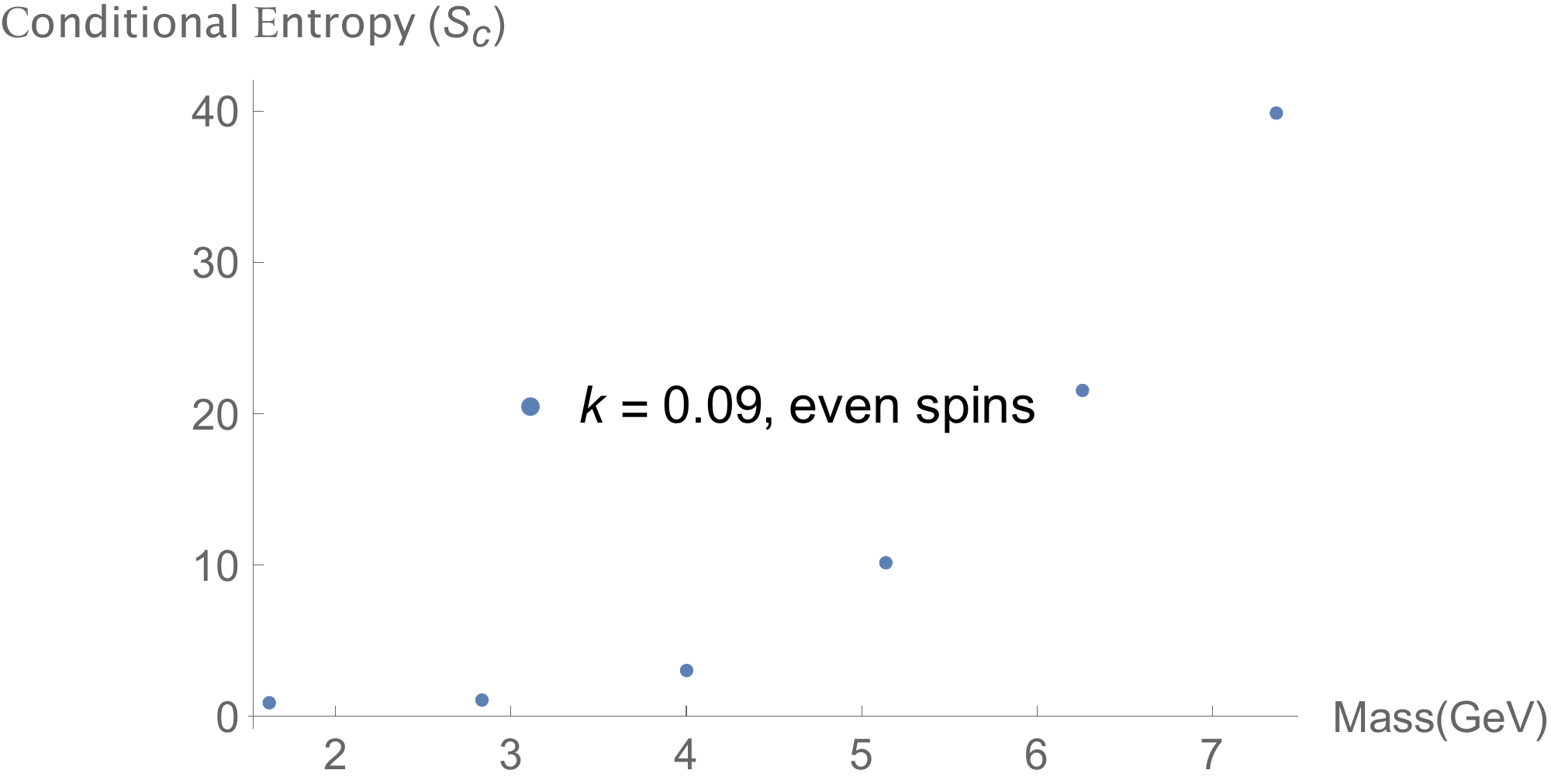}
\label{f9}
\caption{Conditional entropy of glueball states  for $k=0.09$ (line 2 in Table I) for  odd spin (top panel) and even spin  (bottom panel) as a function of the glueball masses.}
\end{figure}
\begin{figure}[H]
\includegraphics[width=8.45cm]{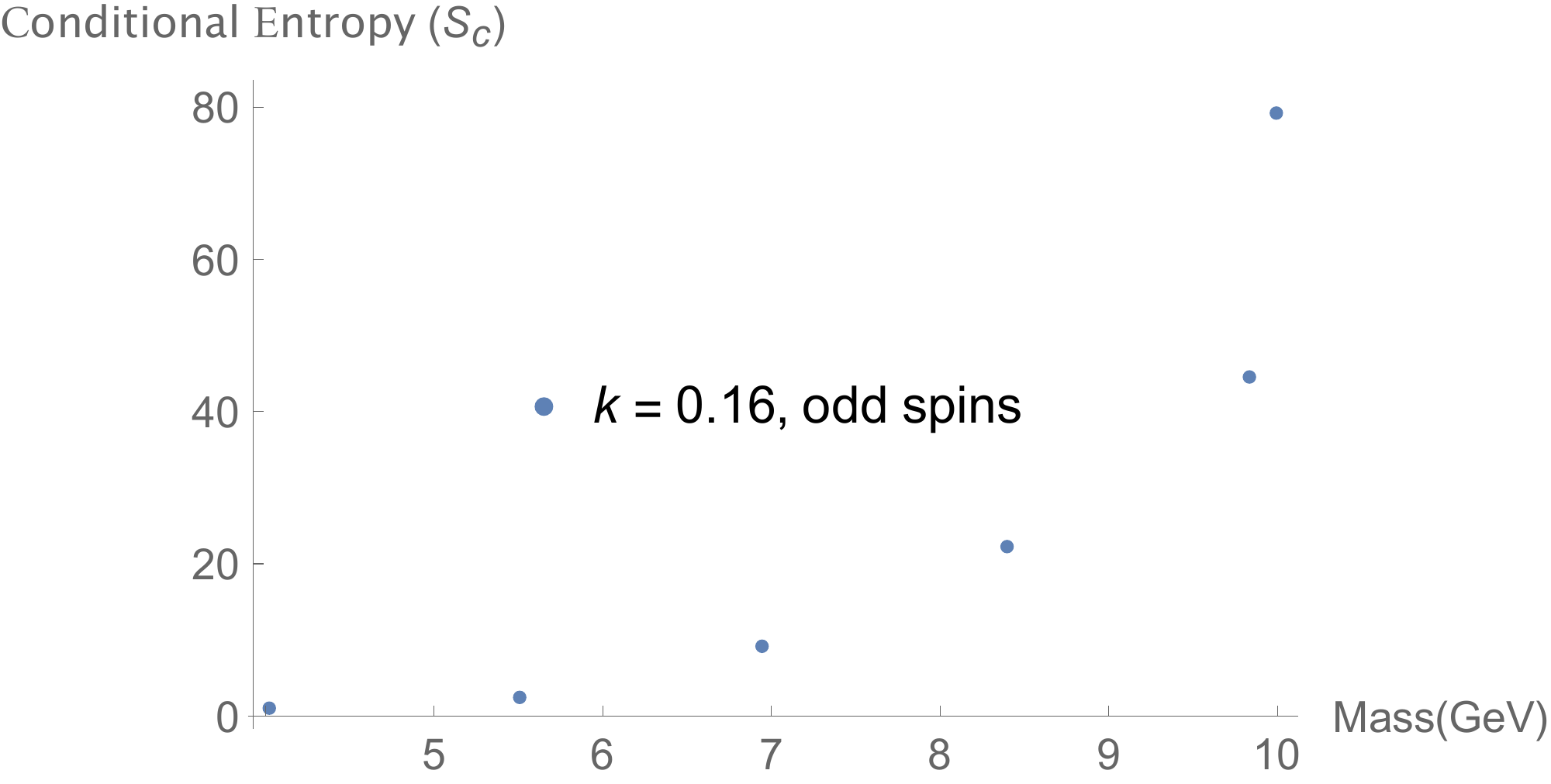}
\includegraphics[width=8.45cm]{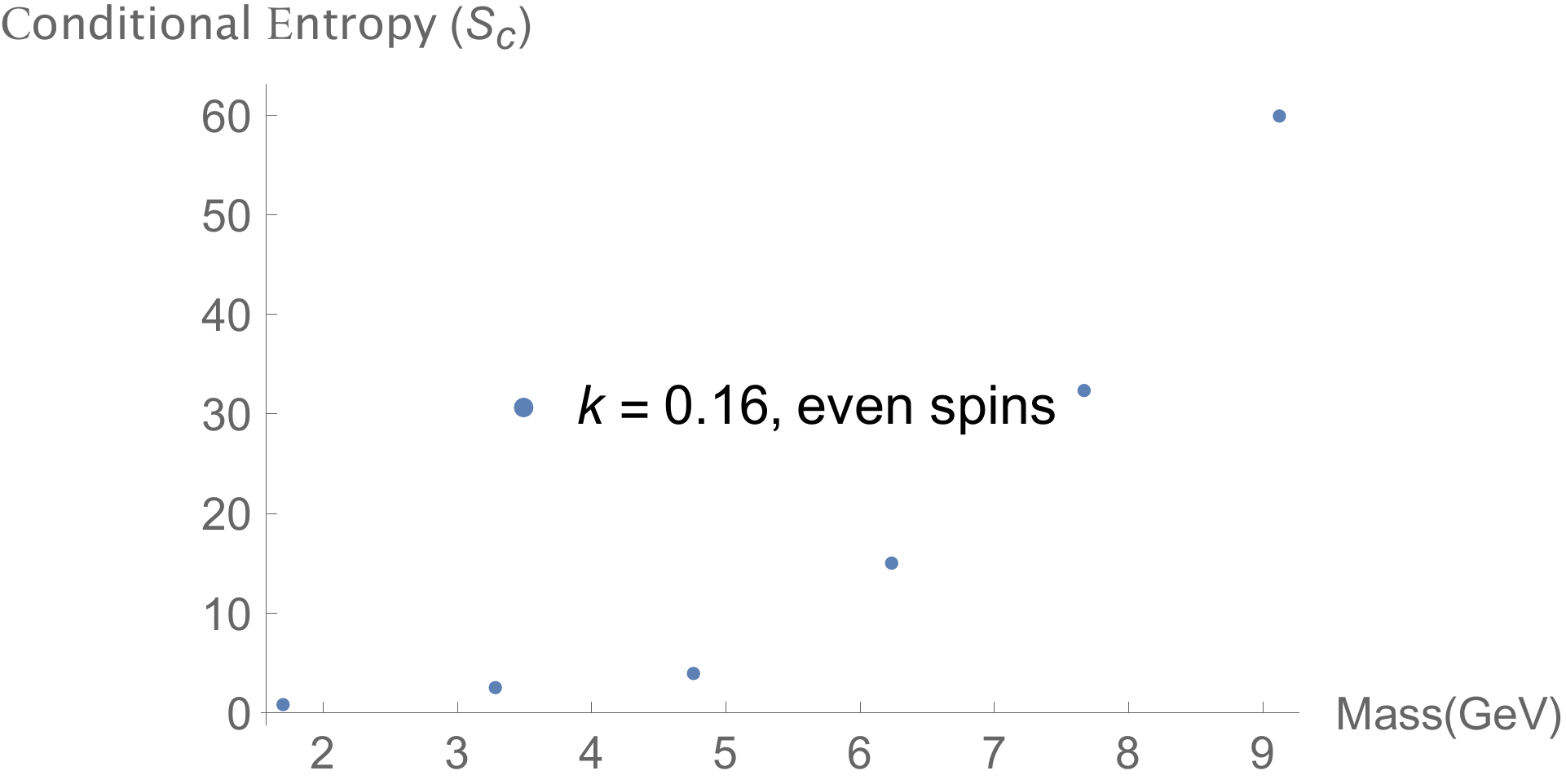}\label{f16}
\caption{Conditional entropy of glueball states  for $k=0.16$ (line 3 in Table I) for  odd spin (top panel) and even spin  (bottom panel) as a function of the glueball masses.}
\end{figure}
Figs. 2, 3 and 4 illustrate the dependence of the conditional entropy with 
the masses of the glueball states, for $k=0.04, k=0.09$ and $k=0.16$, respectively. It is noticeable, as 
in the analysis regarding Fig. 1, that the lower the glueball mass, the lower the conditional entropy is. 
Hence, the conditional entropy is an additional technique 
that can indicate the behavior of glueball states regarding their stability, implying that the states with higher masses are more unstable. 
Moreover,  Fig. 1 shows that for different values of the constant $k$, that defines the quadratic dilaton in Eq. (\ref{qadr}), the higher the value of $k$, the higher the conditional entropy is, for any fixed glueball spin $J$, accordingly. In the next section we present our conclusions. 

\section{Concluding remarks}

Glueballs are not particularly light and have no non-trivial flavor
content. The extraction of a signature in the presence of vacuum fluctuations is therefore more difficult
than for many other hadrons.  
Fig. 1 shows that the higher the glueball states spin $J$, the higher the associated conditional entropy  is. Despite of glueballs to lack  phenomenological support, this study  points towards a manner 
to analyze glueballs stability and production, in the context of lattice AdS/QCD. Figs. 2, 3 and 4 illustrate a quantitative analysis, relating the conditional entropy to the  glueballs masses, for different values of the constant $k$, that defines the quadratic 
dilaton \eqref{qadr}. Irrespectively of the value of $k$ here studied, the 
conditional entropy increases as a function of the glueballs masses. Moreover, 
the conditional entropy is a monotonic increasing function of $k$, for fixed values of the glueball spin, according to Fig. 1. 
This analysis is an useful technique to point toward quantitative physical features of glueball states, 
that still lack in the literature, despite of the advances in lattice QCD. 
 Topological mass constraints could  be further employed, in the 
deformed defects setup, to refine the analysis  presented here \cite{Bernardini:2012bh}.  

\acknowledgments
 
The work of AEB is supported by the Brazilian Agencies FAPESP (grant 2015/05903-4) and CNPq (grant No. 300809/2013-1 and grant No. 440446/2014-7).  NRFB is partially supported by CNPq. RdR is grateful to CNPq (grants No. 303293/2015-2 and No. 473326/2013-2),  to FAPESP (grant 2015/10270-0), for partial financial support, and to R. Casadio, for fruitful discussions.

\end{document}